\documentclass[A4,useAMS,usenatbib,usegraphicx]{mn2e}

\usepackage{amsmath}
\usepackage{amssymb}
\usepackage{times}

\newcommand{\be}{\begin{equation}}
\newcommand{\ee}{\end{equation}}
\newcommand{\ba}{\begin{eqnarray}}
\newcommand{\ea}{\end{eqnarray}}

\newcommand{\WMAP}    {{\sl WMAP}}

\usepackage{aas_macros}
\usepackage{graphicx}
\usepackage{bm}
\usepackage{natbib}
\usepackage{mathrsfs}
\usepackage{color}
\usepackage[mediumspace,Gray,squaren]{SIunits}

\usepackage{hyperref}
\newcommand{\adsurl}[1]{\href{#1}{ADS}}
\providecommand{\url}[1]{\href{#1}{#1}}

\title[Constraints from BAO and LSS]{Cosmological constraints from baryon acoustic oscillations and clustering of large-scale structure}
\author[Addison, Hinshaw \& Halpern]{G. E. Addison$^{1}$\thanks{E-mail: gaddison@phas.ubc.ca}, G. Hinshaw$^{1}$ and M. Halpern$^{1}$\\
$^{1}$Department of Physics and Astronomy, University of British Columbia, 6224 Agricultural Road, Vancouver, BC V6T 1Z1, Canada}

\begin{document}

\date{Accepted xxx. Received xxx; in original form xxx}

\pagerange{\pageref{firstpage}--\pageref{lastpage}} \pubyear{2013}

\maketitle

\label{firstpage}

\begin{abstract}
\noindent
We constrain cosmological parameters using combined measurements of the baryon acoustic oscillation (BAO) feature in the correlation function of galaxies and Ly-$\alpha$ absorbers that together cover $0.1<z<2.4$. The BAO position measurements alone -- without fixing the absolute sound horizon `standard ruler' length with cosmic microwave background (CMB) data -- constrain $\Omega_m=0.303\pm0.040$ (68 per cent confidence) for a flat $\Lambda$CDM model, and $w=-1.06^{+0.33}_{-0.32}$, $\Omega_m=0.292^{+0.045}_{-0.040}$ for a flat $w$CDM model. Adding other large-scale structure (LSS) clustering constraints -- correlation function shape, the Alcock-Paczynski test and growth rate information -- to the BAO considerably tightens constraints ($\Omega_m=0.290\pm0.019$, $H_0=67.5\pm2.8~$km$~$s$^{-1}~$Mpc$^{-1}$, $\sigma_8=0.80\pm0.05$ for $\Lambda$CDM, and $w=-1.14\pm0.19$ for $w$CDM). The LSS data mildly prefer a lower value of $H_0$, and a higher value of $\Omega_m$, than local distance ladder and type IA supernovae (SNe) measurements, respectively. While tension in the combined CMB, SNe and distance ladder data appears to be relieved by allowing $w<-1$, this freedom introduces tension with the LSS $\sigma_8$ constraint from the growth rate of matter fluctuations. The combined constraint on $w$ from CMB, BAO and LSS clustering for a flat $w$CDM model is $w=-1.03\pm0.06$.
\end{abstract}


\section{Introduction}

Recent measurements of temperature anisotropy in the cosmic microwave background (CMB) continue to strongly support the standard $\Lambda$CDM cosmological model, and now constrain its parameters to one or two per cent \citep[e.g.][]{hinshaw/etal:prep,story/etal:prep,sievers/etal:prep,planckparams:prep}. While upcoming polarization data may yield insight into the physics of the early universe, with the recent release of \emph{Planck} cosmology results, future improvements in $\Lambda$CDM constraints from the CMB will be modest. In the coming years, the precision of low-redshift cosmological constraints will significantly increase relative to those from the CMB, predominantly due to intensive efforts to constrain the evolution of dark energy \citep[DE; e.g.][]{albrecht/etal:2006,weinberg/etal:prep}.

One powerful low-redshift probe is the measurement of position of the baryon acoustic oscillation (BAO) feature in the correlation function of large-scale structure (LSS), the imprint of sound waves in the pre-recombination plasma. The BAO measurements are one of the few low-redshift cosmological probes that are limited by statistical, rather than systematic, uncertainties \citep[e.g.][and references therein]{weinberg/etal:prep}.

The BAO scale observed in LSS correlations is intimately related to the photon acoustic scale measured with high precision in the CMB fluctuations \citep{eisenstein/hu/tegmark:1998}. Typically, BAO constraints are discussed, and compared with other low-redshift measurements, using CMB data to fix or tightly constrain portions of parameter space. In this work, we consider an alternative approach, and examine cosmological constraints from the latest BAO (and associated LSS clustering) measurements \emph{without} strong CMB-based priors.

CMB and low-redshift measurements are subject to different observational issues and challenges, and affected differently by, for instance, any non-standard early-universe physics or late-time expansion. A hypothetical deviation from $\Lambda$-acceleration may first appear as a tension between CMB and low-redshift data, or between different low-redshift probes. Such tension in general could also arise from statistical fluctuations, systematic uncertainties that are incorrectly quantified, alternative extensions to the standard model, or some combination of these factors. The ability to disambiguate these possibilities is crucial if we are to get the most out of data from current and future low-redshift experiments. Already, mild tensions have been reported between BAO and local distance ladder measurements of $H_0$ in the context of a $\Lambda$CDM model, when analysed in conjunction with CMB data \citep[e.g.][]{anderson/etal:2012,hou/etal:prep,planckparams:prep}. We will show that there is now evidence for mild BAO--distance ladder tension even without calibrating the BAO measurements using the CMB constraint on the acoustic scale. Similarly, current BAO and LSS precision is now sufficient to permit a meaningful comparison with type IA supernovae (SNe) measurements of the low-redshift expansion rate without strong CMB priors. We show that the BAO and LSS data mildly prefer a higher matter density, $\Omega_m$, than a recent SNe compilation.

The $\Lambda$CDM tensions between the CMB and distance ladder or SNe measurements reported by \emph{Planck} \citep{planckparams:prep} are both effectively relieved by allowing the dark energy equation of state $w<-1$. We will show, however, that reducing $w$ below $-1$ introduces some tension with the amplitude of matter fluctuations, $\sigma_8$, measured using growth rate constraints from redshift-space distortions.

We anticipate that comparing results from different low-redshift probes, both together, and separately from, CMB constraints, will prove increasingly useful as data precision continues to improve.

In Section 2 we discuss the physical information provided by BAO measurements; Section 3 deals with our choice of data and fitting methodology; results are presented in Section 4, and a discussion and conclusions follow in Sections 5 and 6.

\section{Cosmology from the baryon acoustic oscillation scale}

The BAO feature in the LSS correlation function is the consequence of acoustic waves in the pre-recombination baryon -- photon plasma, caused by the opposing forces of gravity and radiation pressure \citep{peebles/yu:1970}. A characteristic scale, roughly the distance these waves have propagated prior to recombination, is imprinted into the matter correlation function when baryons and photons decouple. This scale is typically called the sound horizon and is given by \citep{hu/sugiyama:1996,eisenstein/hu:1998}
\be
r_s=\int_{z_{\rm drag}}^{\infty}\frac{c_s(z)}{H(z)}dz,
\ee
where $z_{\rm drag}$ is the redshift at which baryons ceased to be influenced by Compton drag from photons, $c_s=c/\sqrt{3(1+R)}$ is the sound speed, a function of the ratio of baryon and photon momentum densities, $R=3\rho_b/4\rho_{\gamma}$. In order to be consistent with existing BAO analyses, we adopt the definitions, and fitting formulae for $z_{\rm drag}$, from \cite{eisenstein/hu:1998}. For the cosmologies we consider, rescaling  a numerical calculation of $z_{\rm drag}$ from the CAMB distribution \citep{lewis/etal:2000}, as in \cite{hou/etal:prep} and \cite{planckparams:prep}, changes the inferred BAO constraints by less than 0.2 per cent\footnote{E. Komatsu, priv. comm.} compared to our treatment, which is a negligible difference for current BAO precision.

Studies based on catalogues of LSS tracers measure separations in the radial (line-of-sight; redshift) and transverse (angular) directions. The observables corresponding to the BAO scale are $\Delta z=H(z)\,r_s/c$ and $\Delta\theta=r_s/(1+z)D_A$, respectively, where $D_A$ is the comoving angular diameter distance. Constraints are typically reported in terms of these quantities or $D_V/r_s$, where $D_V$ is a combination of radial and transverse distances \citep{eisenstein/etal:2005}:
\be
D_V(z)=\left[(1+z)^2D_A^2(z)\frac{cz}{H(z)}\right]^{1/3}.
\ee

The large size of the sound horizon, approximately 150 co-moving Mpc, means the BAO feature remains identifiable in low-redshift LSS despite the effects of non-linear growth of structure \citep[e.g.][]{tegmark:1997,eisenstein/seo/white:2007}, however the contrast of the feature is fairly low, essentially because baryons represent a small fraction of the total matter density. Measuring BAO therefore requires surveys that sample large cosmological volumes. The BAO feature in the galaxy correlation function was first detected by the Sloan Digital Sky Survey \citep[SDSS;][]{eisenstein/etal:2005} and Two-degree-Field Galaxy Redshift Survey \citep[2dFGRS;][]{cole/etal:2005}, and has subsequently also been measured using galaxy samples over $0< z<1$ by the Six-degree-Field Galaxy Survey \citep[6dFGS;][]{beutler/etal:2011}, the WiggleZ Dark Energy Survey \citep{blake/etal:2011}, and the Baryon Oscillation Spectroscopic Survey \citep[BOSS;][]{anderson/etal:2012}.

Recently, the BAO feature has been measured in the correlation function of Lyman-$\alpha$ (Ly-$\alpha$) absorbers at $2\lesssim z\lesssim3$ using sight-lines to BOSS quasars \citep{busca/etal:2013,slosar/etal:2013}. The Ly-$\alpha$ measurements are more sensitive to the BAO feature in the radial direction; we follow \cite{busca/etal:2013} in using the constraint on the radial BAO scale relative to a fiducial model,
\be
\alpha_r=\frac{[H(z)\,r_s]_{\rm fid}}{H(z)\,r_s}.
\ee
to constrain cosmological models.

Jointly considering the transverse and radial Ly-$\alpha$ constraints somewhat tightens constraints but does not impact qualitatively on our conclusions. Future Ly-$\alpha$ data may warrant a more detailed approach than we consider here.

\subsection{Constraints from BAO position alone}

Measurements of the BAO scale are sensitive to the low-redshift expansion rate through $D_A$ and $H(z)$, and so constrain $\Omega_m$  (with $\Omega_{\Lambda}$ determined implicitly in a flat $\Lambda$CDM model). A joint fit to BAO data over a range of redshift also constrains an overall normalisation that in this work we will express as the combination $H_0\,r_s$. Going beyond the standard model, BAO measurements alone also constrain parameters that modify the low-redshift expansion rate, such as $w$. BAO-only constraints on $\Omega_m$ and $w$ can be directly compared with those from, for instance, type IA SNe, without requiring an external constraint on the absolute `standard ruler' scale $r_s$ from CMB anisotropy. This comparison is now becoming meaningful owing to the addition of the high-redshift Ly-$\alpha$ BAO constraint to existing $z<1$ BAO measurements.

It should be noted that we view the $w$CDM fits in this work as a mathematical exercise in expansion history parametrization -- something like a $\Lambda$CDM null-test -- and do not present or discuss any particular physical model that could give rise to $w\neq-1$. We also do not investigate allowing additional freedom in $w$, for instance through the popular $w_0-w_a$ parametrization \citep{chevallier/polarski:2001,linder:2003}, since BAO measurements alone do not (yet) usefully constrain such additional parameters.

\subsection{Additional constraints from LSS clustering}

The clustering of LSS tracers contains cosmological constraints beyond simply the position of the BAO feature, although extracting this  information typically requires somewhat stronger assumptions regarding the bias of the tracers or the underlying shape of the matter power spectrum. We here briefly review the additional constraints that we incorporate in the second stage of our analysis.

Marginalising over uncertainties in bias, one can use the entire measured power spectrum as a standard ruler, rather than just the BAO position feature \citep[e.g.][]{eisenstein/etal:2005,sanchez/etal:2008,shoji/etal:2009}. \cite{blake/etal:2011} used this approach to constrain the parameter $A_s=100D_V\sqrt{\Omega_mh^2}/cz$ from the overall power spectrum shape measured from WiggleZ. An additional constraint can be obtained by matching the transverse and radial clustering shapes (i.e. by enforcing statistical isotropy) -- known as the Alcock-Paczynski (AP) test \citep{alcock/paczynski:1979}. This constrains the product of $D_A$ and $H$, which can be expressed as, for example, $F(z)=(1+z)D_A(z)H(z)/c$. It is important to account for anisotropy from redshift-space distortion due to LSS tracer peculiar velocity \citep{kaiser:1987} when applying the AP test \citep{ballinger/peacock/heavens:1996}. Provided non-linearities can be robustly treated, the inclusion of redshift-space power spectrum data constrains $f(z)\sigma_8(z)$, where $f=d\ln\delta/d\ln a$ is the linear growth rate, and $\sigma_8(z)=\sigma_8[\delta(z)/\delta(0)]$, where $\sigma_8$ is the rms linear mass fluctuations in spheres of radius $8~h^{-1}~$Mpc at redshift zero.

The constraints on $\Omega_mh^2$ from the shape of the galaxy power spectrum \citep[e.g.][]{eisenstein/etal:2005,reid/etal:2010,chuang/wang:2013,chuang/etal:2013} mean $H_0$ and $r_s$ can be separately constrained. This is a powerful and useful feature because it allows constraints on $H_0$ from LSS clustering to be compared to other more direct $H_0$ measurements for a given cosmological model (Section 5.1). Similarly, the joint constraints on $\Omega_m$ and $\sigma_8$ can be compared with weak lensing and cluster counting analyses (Section 5.3).

When fitting the growth rate data in a $w$CDM model, we use the following expression for the linear growing mode solution, valid for a flat universe, which is assumed throughout this work  \citep{silveira/waga:1994}
\be
\delta(z)=\frac{1}{1+z}\,_2F_1\left[-\frac{1}{3w},\frac{w-1}{2w},1-\frac{5}{6w},-(1+z)^{3w}\frac{1-\Omega_m}{\Omega_m}\right],
\ee
where $_2F_1$ is the hypergeometric function.

\section{Data and model fitting}

\begin{table*}
  \centering
  \caption{Measurements of BAO position and large-scale clustering (LSS) clustering used in this analysis. The LSS clustering measurements include constraints on the acoustic parameter, $A_s$, the Alcock-Paczynski (AP) parameter, $F$, and the growth rate, $f\sigma_8$, as defined in Section 2. Note that, while some BAO constraints are included in the LSS data set, the BAO-only data are not a subset of the LSS data (see Section 4.1). Cosmological constraints from the BAO and LSS data are discussed in Sections 4 and 5.}
  \begin{tabular}{lllll}
\hline
Label&Survey&$z_{\rm eff}$&Constraint&Reference\\
\\
\hline
BAO-only&6dFGS&0.106&$r_s/D_V=0.336\pm0.015$&\cite{beutler/etal:2011}\\
\\
&SDSS, DR7$^a$&0.35&$D_A/r_s=6.875\pm0.246$&\cite{xu/etal:2013}\\
&&0.35&$Hr_s$=$(12895\pm1070)~$km$~$s$^{-1}$&\\
\\
&WiggleZ$^b$&0.44&$r_s/D_V=0.0916\pm0.0071$&\cite{blake/etal:2011}\\
&&0.60&$r_s/D_V=0.0726\pm0.0034$&\\
&&0.73&$r_s/D_V=0.0592\pm0.0032$&\\
\\
&BOSS, DR9 CMASS$^c$&0.57&$D_A(r_s^{\rm fid}/r_s)=(1408\pm45)~$Mpc&\cite{anderson/etal:prep}\\
&&0.57&$H(r_s/r_s^{\rm fid})=(92.9\pm7.8)~$km$~$s$^{-1}~$Mpc$^{-1}$&\\
\\
&BOSS, Ly-$\alpha$ forest$^d$&2.4&$\alpha_r$; see text&\cite{slosar/etal:2013}\\
\hline

LSS&6dFGS&0.106&$r_s/D_V=0.336\pm0.015$&\cite{beutler/etal:2011}\\
\\
&SDSS, DR7$^a$&0.35&$D_A/r_s=6.875\pm0.246$&\cite{xu/etal:2013}\\
&&0.35&$Hr_s$=$(12895\pm1070)~$km$~$s$^{-1}$&\\
\\
&WiggleZ$^b$&0.44&$A_s=0.474\pm0.034$&\cite{blake/etal:2012}\\
&&0.60&$A_s=0.442\pm0.020$&\\
&&0.73&$A_s=0.424\pm0.021$&\\
&&0.44&$F=0.482\pm0.049$&\\
&&0.60&$F=0.650\pm0.053$&\\
&&0.73&$F=0.865\pm0.073$&\\
&&0.44&$f\sigma_8=0.413\pm0.080$&\\
&&0.60&$f\sigma_8=0.390\pm0.063$&\\
&&0.73&$f\sigma_8=0.437\pm0.072$&\\
\\
&BOSS, DR9 CMASS$^e$&0.57&$H=(87.6\pm7.2)~$km$~$s$^{-1}~$Mpc$^{-1}$&\cite{chuang/etal:2013}\\
&&0.57&$D_A=(1396\pm74)~$Mpc&\\
&&0.57&$\Omega_mh^2=0.126\pm0.019$&\\
&&0.57&$f\sigma_8=0.428\pm0.069$&\\
\\
&BOSS, Ly-$\alpha$ forest$^d$&2.4&$\alpha_r$; see text&\cite{slosar/etal:2013}\\
\hline
\end{tabular}
\raggedright{
\newline
$^a$ the SDSS DR7 post-reconstruction measurements of $D_A/r_s$ and $Hr_s$ are correlated with correlation coefficient 0.57\\
$^b$ the covariance matrix for the WiggleZ measurements is given in equation (4) of \cite{hinshaw/etal:prep} for $D_V/r_s$ and Table 2 of \cite{blake/etal:2012} for $A_s$, $F$ and $f\sigma_8$\\
$^c$ the BOSS DR9 CMASS sample post-reconstruction measurements of $D_A(r_s/r_s^{\rm fid})$ and $H(r_s/r_s^{\rm fid})$ are correlated with correlation coefficient 0.55; the fiducial sound horizon, $r_s^{\rm fid}$, adopted by \cite{anderson/etal:prep} is $r_s^{\rm fid}=153.19~$Mpc\\
$^d$ the fiducial value of $H\,r_s$ (equation 3) adopted for the BOSS Ly-$\alpha$ analysis is $3.62\times10^4~$km$~$s$^{-1}$\\
$^e$ the covariance matrix between parameters in the full BOSS CMASS clustering analysis is given in equation (26) of \cite{chuang/etal:2013}}
\end{table*}

The BAO data used in our analysis are listed in Table 1. We opt to use recent analyses of the SDSS Data Release (DR) 7 and BOSS DR9 CMASS samples that constrain the BAO position in both the radial and transverse directions \citep{xu/etal:2013,anderson/etal:prep}. We also use SDSS and CMASS constraints from analyses that attempt reconstruction of the linear density field \citep{eisenstein/etal:2007}. Earlier results \citep[e.g.][]{eisenstein/etal:2005,percival/etal:2010} are in good agreement with these newer analyses.

The bottom part of Table 1 lists constraints used in our expanded analysis, including correlation function shape, AP and growth rate measurements, in addition to BAO feature position. It should be noted that the SDSS and BOSS CMASS samples have been re-analyzed multiple times \citep[e.g.][]{reid/etal:2010,samushia/etal:2011,padmanabhan/etal:2012,anderson/etal:2012,sanchez/etal:2013}. Some choice of which constraints to adopt must therefore be made. Ideally, one would jointly fit to the `raw' data -- the correlation function or power spectrum measurements -- from each sample, rather than to derived quantities such as $D_V/r_s$. The latter approach is, however, adequate for this work, since our conclusions are largely qualitative and relatively insensitive to the exact choice of constraint combination (although see comments on goodness-of-fit in Section 4.1).

\cite{chuang/etal:2013} present a `single probe' analysis of the BOSS CMASS data, jointly using BAO position, clustering shape and redshift-space distortion information to extract cosmological constraints, independent of dark energy evolution model, with broad priors on parameters not well-constrained from their data. This work built upon earlier analysis using the SDSS sample at $z_{\rm eff}=0.35$ \citep[e.g.][]{chuang/wang:2013}. We opt not to include constraints from the earlier work in all our fits since they were found to be somewhat more sensitive to, for example, the range of separation scales used in correlation function fitting, although we consider the effect of their inclusion in the context of LSS constraints on $H_0$ in Section 5.1.

The interdependence between BAO position and growth rate \citep{beutler/etal:2012} has not been quantified for the 6dFGS and so for the expanded LSS clustering analysis we choose to use only the BAO scale constraint, and similarly for the SDSS $z_{\rm eff}=0.35$ sample. We do not include growth rate constraints from surveys outside those used for the BAO-only fit. This is a slightly arbitrary choice, however the growth rate constraints from other surveys, including 2dFGRS \citep{hawkins/etal:2003}, 2SLAQ \citep{ross/etal:2007}, VVDS \citep{guzzo/etal:2008}, and VIPERS \citep{delatorre/etal:2013}, are statistically consistent with the constraints used here, and do not provide significant improvements for the models considered.

The WiggleZ and 6dFGS analyses fixed several parameters, including the baryon density and primordial power spectrum index, $n_s$, based on \emph{Wilkinson Microwave Anisotropy Probe} (\WMAP) constraints \citep{komatsu/etal:2011}. We do not view these assumptions as likely to lead to either bias in inferred BAO constraints, or significant interdependence between CMB and BAO or LSS constraints for the $\Lambda$CDM and $w$CDM models discussed in this work. The WiggleZ and 6dFGS data sets constrain $D_V/r_s$ or $A_s$ to around 5 per cent. The sound horizon varies very weakly with $\Omega_bh^2$ for the cosmologies we consider ($r_s\propto(\Omega_bh^2)^{-0.13}$), and is, furthermore, constrained to better than 1.5 per cent by the CMB \citep{planckparams:prep}. \cite{eisenstein/etal:2005} found that the acoustic parameter, $A_s$, scales as $n_s^{-0.35}$ from fitting to SDSS data. Assuming this relation, shifting a WiggleZ $A_s$ value by $0.5\sigma$ requires a $7\sigma$ shift in $n_s$ (as inferred from current CMB data for either $\Lambda$CDM or $w$CDM). The high precision of CMB constraints essentially means that uncertainty in these parameters is subdominant to sample variance for the WiggleZ and 6dFGS data. Far weaker CMB-based assumptions can be, and are, adopted for the more precise BAO and LSS measurements from SDSS and BOSS.

\cite{blake/etal:2011c} used clustering information on scales down to $k=0.3~h$Mpc$^{-1}$ to extract the WiggleZ growth rate constraints. To check that uncertainties relating to non-linear effects on these scales are not significantly biasing our results, we repeated the BAO plus LSS fit with the WiggleZ constraints removed. While parameter constraints are degraded (by up to a factor of two in the case of $\sigma_8$), the shifts in the peaks of the marginalised parameter posterior probability distributions are small -- at most around 20 per cent of a statistical standard deviation -- for all parameters.

In order to fit cosmological parameters, we introduce a likelihood function
\be
-2\ln\mathcal{L}=({\bf x}-{\bf d})^T{\bf C}^{-1}({\bf x}-{\bf d})-2\ln\mathcal{L}_{\textrm{Ly}\alpha},
\ee
where ${\bf x}$, ${\bf d}$ and ${\bf C}$ are the model predictions, mean data values and data covariance matrix of the 6dFGS, SDSS, WiggleZ and BOSS CMASS constraints (including covariance between different constraints from the same survey).

\cite{slosar/etal:2013} report a non-Gaussian and asymmetric probability distribution for $\alpha_r$ from the BOSS Ly-$\alpha$ forest measurements. We add the `top-hat' systematic uncertainty of $\pm0.02$ obtained by \cite{slosar/etal:2013} based on the scatter between different fitting methods linearly to the statistical uncertainty estimates to give $\alpha_r=0.987^{+0.055\,+0.096\,+0.143}_{-0.053\,-0.087\,-0.122}$ ($\pm1$, 2 and 3$\sigma$ errors). We then construct a polynomial likelihood of the form
\be
-2\ln\mathcal{L}_{\textrm{Ly}\alpha} = \left\{
         \begin{array}
         {l@{\quad {\rm for}\quad}l}
 C_1x^2+D_1x^3+E_1x^4 & \alpha_r\leq0.987 \\
C_2x^2+D_2x^3+E_2x^4 & \alpha_r>0.987,
 \end{array}
\right.  \ee 
where $x=|\alpha_r^{\rm model}-0.987|$, and the constants $C_{1,2}$, $D_{1,2}$ and $E_{1,2}$ are determined by matching to the $\pm1$, 2 and 3$\sigma$ uncertainties in $\alpha_r$. Our analysis is insensitive to exactly how the Ly-$\alpha$ likelihood is treated -- assuming a purely Gaussian likelihood, for instance, does not have any significant impact on our results.

\cite{busca/etal:2013} present an analysis of largely the same Ly-$\alpha$ sample used by \cite{slosar/etal:2013}, but with various differences in data cuts and methodology \citep[see Section 5.4 of][]{slosar/etal:2013}. We discuss the effects of using the \cite{busca/etal:2013} results rather than those of \cite{slosar/etal:2013} in Section 5.1.

We explore parameter spaces using a Markov Chain Monte Carlo (MCMC) method, specifically the affine-invariant `stretch step' ensemble sampler proposed by \cite{goodman/weare:2010} and parallelised in the emcee\footnote{http://dan.iel.fm/emcee/} Python module by \cite{foreman-mackey/etal:2013}. For the relatively small number of parameters we consider here, a tuned Metropolis sampler \citep{metropolis/etal:1953,dunkley/etal:2005} would likely be more efficient, however we expect that the flexibility and parallel nature of the stretch step approach \citep[above references and][]{akeret/etal:prep} to prove useful as we include more data sets in future analysis.

\section{Results}

In this section we discuss constraints from the BAO-only and LSS clustering fits and compare to the latest CMB results. Here and throughout, `CMB' refers to a joint fit to the \emph{Planck} 2013 temperature and lensing power spectra \citep{planckspectrum:prep,plancklensing:prep}, \WMAP\, 9-year polarization data \citep{bennett/etal:prep2} and small-scale temperature power spectra from the Atacama Cosmology Telescope \citep[ACT;][]{das/etal:prep} and the South Pole Telescope \citep[SPT;][]{reichardt/etal:2012}. We show constraints from MCMC chains provided by the \emph{Planck} collaboration\footnote{http://pla.esac.esa.int/pla/aio/planckResults.jsp?}. Comparisons with other data sets are made in Section 5.

\subsection{$\Lambda$CDM}

\begin{table*}
  \centering
  \caption{Cosmological constraints from BAO position only (using constraints from top part of Table 1), BAO position plus a baryon density prior from estimation of the primordial deuterium abundance (Section 5.1), LSS clustering (bottom part of Table 1), and the latest CMB measurements \citep[\emph{Planck} temperature and lensing power spectra plus \WMAP\, polarization and ACT and SPT small-scale temperature power spectra;][]{planckparams:prep}. We show 68.3 per cent confidence intervals for all parameters and additionally show 95.5 per cent confidence intervals for $w$.}
  \begin{tabular}{llcccc}
Model&Parameter&BAO-only&BAO+$\Omega_bh^2$&LSS&CMB-only\\
\hline
$\Lambda$CDM&$\Omega_m$&$0.304\pm0.040$&$0.308^{+0.040}_{-0.041}$&$0.290\pm0.019$&$0.307\pm0.013$\\
&$H_0\,r_s/\,10^4~$km$~$s$^{-1}$&$1.035\pm0.024$&$1.033\pm0.024$&$1.048\pm0.022$&$1.027^{+0.018}_{-0.017}$\\
&$H_0/\,$km$~$s$^{-1}~$Mpc$^{-1}$&--&$68.9\pm3.0$&$67.5\pm2.8$&$67.9\pm1.0$\\
&$\sigma_8$&--&--&$0.802\pm0.047$&$0.8233\pm0.0097$\\
\hline
$w$CDM&$w$&$-1.06^{+0.33\,+0.63}_{-0.32\,-0.72}$&$-1.11^{+0.32\,+0.61}_{-0.32\,-0.71}$&$-1.14^{+0.19\,+0.36}_{-0.19\,-0.42}$&$-1.49^{+0.30\,+0.62}_{-0.29\,-0.42}$\\
&$\Omega_m$&$0.292^{+0.045}_{-0.040}$&$0.299^{+0.042}_{-0.040}$&$0.275^{+0.030}_{-0.029}$&$0.204^{+0.052}_{-0.051}$\\
&$H_0\,r_s/\,10^4~$km$~$s$^{-1}$&$1.045\pm0.058$&$1.050\pm0.057$&$1.076\pm0.045$&$1.287^{+0.219}_{-0.166}$\\
&$H_0/\,$km$~$s$^{-1}~$Mpc$^{-1}$&--&$70.7\pm7.7$&$70.1\pm4.8$&$85.0\pm10.9$\\
&$\sigma_8$&--&--&$0.77\pm0.07$&$0.96\pm0.08$\\
\hline
\end{tabular}
\end{table*}

For the $\Lambda$CDM fit to BAO position measurements alone we find $\Omega_m= 0.304\pm0.040$ and $H_0\,r_s=(1.035\pm0.024)\times10^4~$km$~$s$^{-1}$. We report mean parameter values and the boundaries of the symmetric 68.3 per cent and, in some cases, 95.5 per cent, confidence intervals. Constraints in the $H_0\,r_s-\Omega_m$ plane are shown in Figure 1 for the combined BAO data as well as the $z=0.106$ 6dFGS, $z=0.57$ BOSS CMASS and $z=2.4$ BOSS Ly-$\alpha$ constraints individually. We show CMB constraints for the $\Lambda$CDM model in the same plane.

Our $\Lambda$CDM parameter space for the expanded LSS clustering fit is spanned by
\be
 \{\Omega_m,H_0,r_s,\sigma_8\}.
 \ee
 For the $\Lambda$CDM model, there is striking agreement for all parameters between the BAO-only, LSS, and CMB constraints (Table 2), consistent with the discussion in Section 5.2 of \cite{planckparams:prep}. Adding the LSS clustering information tightens constraints on $\Omega_m$ by a factor of two over the BAO position data alone, and constrains $H_0$ to around 4 per cent and $\sigma_8$ to around 6 per cent.

Notice that the LSS data contour in Figure 1 contains regions disfavoured in the BAO-only fit. This is because the strongest LSS constraints do not depend explicitly on the sound horizon, $r_s$. The BOSS CMASS and WiggleZ analyses used in the LSS fit  \citep{blake/etal:2012,chuang/etal:2013} focus on the shape of the galaxy correlation function -- the dependence on $r_s$ is effectively marginalised over in the case of \cite{chuang/etal:2013}, and negligible in the case of \cite{blake/etal:2012}, because the BAO feature is not significantly detected in the two-dimensional WiggleZ power spectrum. In the joint fit with the other galaxy survey constraints, the reduction in information about $r_s$ compared to the BAO-only BOSS and WiggleZ analyses leads to a broadening of constraints in the direction roughly perpendicular to the BAO-only BOSS contours (Figure 1). Using the \cite{blake/etal:2012} and \cite{chuang/etal:2013} results does, however, lead to tighter constraints on the parameters of most interest in this analysis -- $\Omega_m$ and $w$, when it is free -- partially breaking the degeneracy apparent in the BAO-only contours.

%Notice that the BAO plus LSS data contour in Figure 1 contains regions disfavoured in the BAO-only fit. The analyses of \cite{blake/etal:2012} and \cite{chuang/etal:2013} apply the AP test to the WiggleZ and BOSS CMASS data, requiring the transverse and radial correlation function shapes to match, and yielding constraints on $D_A$ and $H$ that do not depend on $r_s$. The reduction in information relating to $r_s$ degrades constraints in the direction roughly perpendicular to the CMASS BAO-only contours in the $H_0\,r_s-\Omega_m$ plane, compared to the BAO-only fit. The addition of the AP test results does, however, yield tighter constraints on the parameters of most interest in this analysis -- $\Omega_m$ and $w$, when it is free -- partially breaking the degeneracy apparent in the BAO-only contours of Figure 1.

%Notice that the BAO plus LSS data contour in Figure 1 contains regions disfavoured in the BAO-only fit. This is largely because the BOSS CMASS constraint on $D_A(0.57)$ from the joint analysis of \cite{chuang/etal:2013} is around 50 per cent weaker than the constraint on $D_A(0.57)/r_s$ from the same sample. For the BAO plus LSS fit, we opt to adopt the constraints on $D_A$ and $H$, which yield tighter constraints on $H_0$ and $\Omega_m$, but weaker constraints in the direction roughly perpendicular to the CMASS BAO-only contours in the $H_0\,r_s-\Omega_m$ plane.

The LSS $\Lambda$CDM constraint of $\Omega_m=0.290\pm0.019$ is around 50 per cent weaker than that from current CMB data; it is worth pointing out, though, that it is comparable or stronger than any pre-\emph{Planck} CMB measurement -- the \emph{WMAP-9} 68 per cent uncertainty on $\Omega_m$ is $0.025$, while that from combining \WMAP-9 with SPT data is $0.019$ \citep{calabrese/etal:2013}, for example. It is realistic to expect LSS to constrain $\Omega_m$ with comparable precision to \emph{Planck} in the fairly near future.

The $\chi^2$ value corresponding to the maximum likelihood from the BAO-only $\Lambda$CDM chain is $1.73$. We are fitting to a total of 9 data points with two parameters ($\Omega_m$ and $H_0\,r_s$); the probability for $\chi^2$ per degree of freedom (dof) to exceed the measured value (PTE) is very high -- 0.97. This issue is exacerbated in the fit including the additional LSS data: we find $\chi^2/\textrm{dof}=3.5/(17-4)$ (PTE of 0.995). The high PTE values remain if we remove the Ly-$\alpha$ constraint and our associated non-Gaussian likelihood. Our analysis neglects interdependence between constraints from different surveys. There is partial overlap in both redshift and sky coverage for WiggleZ and BOSS (see Figure 1 of \citealt{drinkwater/etal:2010} and Figure 1 of \citealt{ahn/etal:2012}), however 70\% of the BOSS sky coverage lies outside the WiggleZ regions, and these surveys use somewhat different galaxy selection criteria. We have confirmed that the high PTE remains even if we repeat our fits omitting the constraints from one of these two surveys.

It seems plausible that uncertainties on derived quantities used in this work, such as $D_V/r_s$ and $f\sigma_8$, may be overestimated as a result of individual analyses conservatively choosing methodology approaches that lead to the broadest constraints. \cite{chuang/etal:2012} found that estimating LSS clustering bandpower covariance using lognormal realizations to approximate non-linearities in the density field \citep{coles/jones:1991,percival/etal:2004}, rather than $N$-body simulations, leads to systematically larger uncertainties. This could contribute to the high PTE in the case of the WiggleZ and 6dFGS data, where the lognormal method was used \citep{blake/etal:2011b,beutler/etal:2011}. We therefore expect the issue of high PTE to be largely resolved through a combination of fitting directly to correlation function or power spectrum measurements, and future improvements in LSS analysis methodology.

%Even for surveys with significant redshift overlap (such as WiggleZ and BOSS), the sky coverage and survey windows are very different, as are the galaxy selection criteria. We consequently see no reason to expect significantly correlated sample variance-like uncertainties.

\begin{figure}
	\centering
	\includegraphics[]{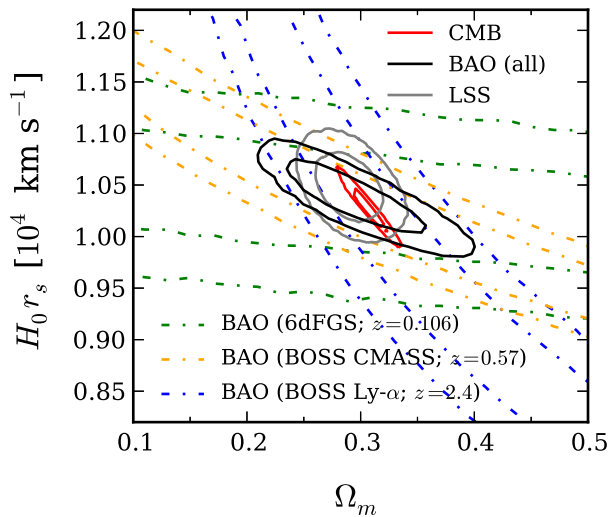}
	\caption{$\Lambda$CDM constraints from measurements of BAO position. We show 68.3 and 95.5 per cent confidence contours for the combined BAO data set (Table 1, top), and for the 6dFGS \citep{beutler/etal:2011}, BOSS CMASS \citep{anderson/etal:prep} and BOSS Ly-$\alpha$ \citep{slosar/etal:2013} data separately, to illustrate their complementarity. Constraints from our expanded analysis including other large-scale structure clustering constraints (Table 1, bottom), and CMB constraints from combining \emph{Planck}, \WMAP, ACT and SPT data \citep{planckparams:prep} are also shown. The BAO, LSS and CMB constraints are in good agreement for the $\Lambda$CDM model.}
\end{figure}

\subsection{$w$CDM}

For $w$CDM, the BAO position measurements alone constrain $w=-1.06^{+0.33}_{-0.32}$, $\Omega_m=0.292^{+0.045}_{-0.040}$, and $H_0\,r_s=(1.045\pm0.058)\times10^4~$km$~$s$^{-1}$. Addition of the clustering shape, AP and growth rate constraints tighten $w$CDM constraints, although not to the extent that they did for $\Lambda$CDM, with the $w$ constraint improving to $-1.14\pm0.19$. The $\chi^2$ for the LSS $w$CDM fit is improved by 0.3 compared to $\Lambda$CDM; currently, there is no significant evidence for departures from $\Lambda$-acceleration from LSS data. The BAO and LSS data constraints on other parameters are fairly robust to allowing freedom in $w$, while the CMB constraints are significantly broadened. The CMB data exhibit a mild ($\sim1.5\sigma$) preference for $w<-1$, which leads to shifts in the other parameters shown in Table 2, since these parameters are highly correlated with $w$ when constrained with the CMB alone \citep[see Figure 21 of][]{planckparams:prep}.

\cite{chuang/etal:2013} found that including growth rate measurements improved constraints on $w$ in a joint fit using the BOSS CMASS sample and CMB data. It is worth noting here that, in our fit to the LSS clustering data only, the growth rate measurements effectively only constrain $\sigma_8$, and contribute minimally to constraints on $\Omega_m$ or $w$. This is because $f\sigma_8$ depends weakly on redshift for $z<1$ and so there is little leverage for current growth rate measurements  to improve expansion history constraints in the absence of an external constraint on $\sigma_8$. In Section 5.2, we show that the growth rate $\sigma_8$ constraint does, however, play an important role when assessing the extent to which allowing $w\neq-1$ can relieve tension between the CMB and low-redshift data sets.

\section{Discussion}

\subsection{Comparison with distance ladder $H_0$ measurements}

Figure 2 compares marginalized constraints on $H_0$. We show the local distance ladder measurements of $H_0=(73.8\pm2.4)~$km$~$s$^{-1}~$Mpc$^{-1}$ measured using Cepheid variable stars and low-redshift type IA SNe observed with the \emph{Hubble Space Telescope} (HST) by \cite{riess/etal:2011}. Similar results were obtained from a more recent re-analysis of HST Cepheid and SNe data using a new estimate of the distance to the Large Magellanic Cloud calibrated using $3.6~\mu$m observations \citep{freedman/etal:2012}.

The BAO position measurements \emph{alone} do not provide any $H_0$ constraint, being sensitive only to the combination $H_0\,r_s$. In addition to $H_0$ and $\Omega_m$, the sound horizon depends on the physical baryon density, $\Omega_bh^2$, though only weakly -- $r_s\propto(\Omega_bh^2)^{-0.13}$ for the cosmological models considered here. We are therefore able to obtain constraints on $H_0$ from the BAO position measurements with the addition of a prior on the baryon density \citep[and the CMB mean temperature, which determines the energy density in radiation, and which we hold fixed to $2.72548~$K;][]{fixsen:2009}. The most precise constraints on the baryon density outside the CMB come from estimates of the primordial deuterium abundance from metal-poor damped Ly-$\alpha$ systems. Recently, \cite{pettini/cooke:2012} found $\Omega_bh^2=0.0223\pm0.0009$ (assuming no non-standard relativistic species) for a system particularly well-suited to this measurement at $z\sim3$. Adopting this constraint as a Gaussian prior and repeating the fit to the BAO data in the top part of Table 1 using the parameter set $\{\Omega_m,\Omega_bh^2,H_0\}$, yields $H_0=68.9\pm3.0~$km$~$s$^{-1}~$Mpc$^{-1}$. Alternatively, using a weak CMB-based prior of $\Omega_bh^2=0.02218\pm0.00130$, five times wider than the CMB-only 68 per cent confidence constraint from \cite{planckparams:prep}, gives $H_0=68.8\pm3.1~$km$~$s$^{-1}~$Mpc$^{-1}$. These values are in good agreement with the LSS clustering constraint and lower by around $1.3\sigma$ than the \cite{riess/etal:2011} value. Note that the baryon density is determined from the CMB power spectrum largely through the relative heights of acoustic peaks, rather than their spacing -- this weak CMB-based baryon density prior is highly robust to modifications that could alter the acoustic scale, such as extra relativistic species.

It should be noted that the Ly-$\alpha$ data dominate the $H_0$ constraint in the BAO position plus baryon density fit; removing this data point, we find $H_0=70.1\pm7.1~$km$~$s$^{-1}~$Mpc$^{-1}$ for the deuterium abundance $\Omega_bh^2$ prior. The high-redshift information partially breaks degeneracies present in the low-redshift constraints (Figure 1). Using the \cite{busca/etal:2013} Ly-$\alpha$ constraint of $\alpha_r^{-1}=0.954\pm0.077$ (for their method 2 and broadband parametrization of their equation 24), rather than that of \cite{slosar/etal:2013}, gives $H_0=66.8\pm3.7~$km$~$s$^{-1}~$Mpc$^{-1}$. On the other hand, using the isotropic Ly-$\alpha$ constraint from \cite{slosar/etal:2013}, rather than simply the measurement in the radial direction, and assuming the isotropic distortion parameter $\alpha_{\rm iso}\propto D_A^{0.2}H^{-0.8}$ \citep{busca/etal:2013}, we find $H_0=69.2\pm2.3~$km$~$s$^{-1}~$Mpc$^{-1}$.

In the expanded LSS clustering analysis, $H_0$ is constrained from the shape of the galaxy correlation function, which is sensitive to $\Omega_mh^2$. The resulting $\Lambda$CDM constraint of $67.5\pm2.8~$km$~$s$^{-1}~$Mpc$^{-1}$ is in mild ($1.7\sigma$) tension with the direct $H_0$ measurement. The choice of which galaxy clustering constraints to rely on has some effect here; adding the constraint on $\Omega_mh^2$ from the $z_{\rm eff}=0.35$ SDSS sample from \cite{chuang/wang:2013}, which was not included in our base LSS fit (Section 3), shifts the $H_0$ constraint to $66.5\pm 2.1~$km$~$s$^{-1}~$Mpc$^{-1}$, which is in tension with the \cite{riess/etal:2011} measurement at around the $2.2\sigma$ level, while remaining in good agreement with the CMB value.

Overall, then, we find that the LSS clustering data exhibit qualitatively the same tension with the distance ladder $H_0$ measurements, assuming a $\Lambda$CDM model, as the \emph{Planck} data, preferring a lower $H_0$ value, although only at fairly weak statistical significance. Clearly, this difference could be largely, or solely, the result of statistical fluctuation. We also discuss the extent to which allowing $w\neq-1$ relieves this, and other, $\Lambda$CDM tensions in Section 5.2, below. Discussion of other extensions, such as increasing the number of effective neutrino species, is deferred to future work.

Due to the weak dependence of the sound horizon on the baryon density, the uncertainties given in this section are largely limited by statistical uncertainties in the BAO and LSS clustering data. There is therefore scope for considerable improvement in (model-dependent) $H_0$ constraints with future LSS clustering data, particularly at high redshift, with minimal dependence on the CMB measurement of the acoustic scale.

\begin{figure}
	\centering
	\includegraphics[]{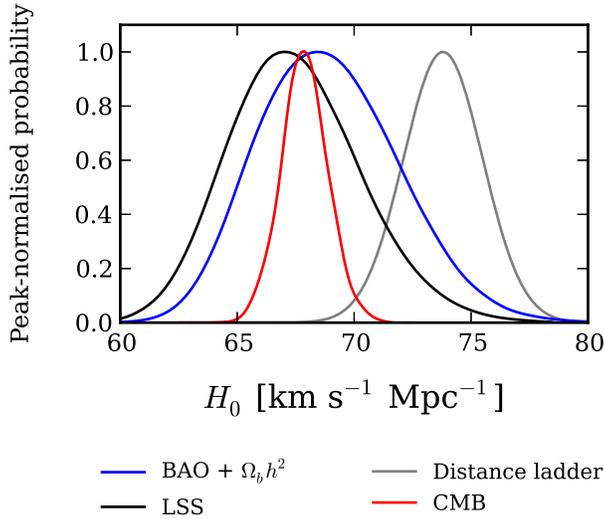}
	\caption{Marginalized $H_0$ constraints for the $\Lambda$CDM model. While BAO position measurements alone do not constraint $H_0$, a constraint may be obtained by either adding information from the shape of the large-scale structure correlation function, or adding a prior on the baryon density \protect\citep[here we use the CMB-independent constraint from primordial deuterium abundance estimated by][]{pettini/cooke:2012}. In either case, there is a mild preference for a lower $H_0$ value than the distance ladder measurements, consistent with recent \emph{Planck} results. Note that the choice of large-scale structure constraints moderately affects this comparison (see text).}
\end{figure}

\subsection{Comparison with type IA supernovae measurements and allowing $w\neq-1$}

Measurements of type IA supernovae brightness as a function of redshift constrain the expansion history, and thus $\Omega_m$ and $w$, if an empirical correlation between luminosity and light curve shape is assumed \citep[for discussion of the role of SNe as DE probes, see][and references therein]{weinberg/etal:prep}. These constraints can be directly compared to those from the BAO position and LSS clustering data. In this work, we compare to the Supernova Legacy Survey (SNLS) compilation, consisting of 472 SNe from various surveys, and analysed by \cite{conley/etal:2011}, with marginalization over nuisance parameters relating to known SNe systematic uncertainties.

For the $\Lambda$CDM model, the SNLS compilation gives $\Omega_m=0.232\pm0.039$ (68 per cent confidence). This is in agreement with, though around $1.3\sigma$ lower than, the BAO or LSS constraints of $0.304\pm0.040$ and $0.290\pm0.019$, respectively. As with $H_0$, BAO and LSS constraints on $\Omega_m$ are expected to improve significantly in the relatively near future (Section 5.4).

It is interesting to consider the combined LSS, SNe, distance ladder and CMB data in the context of the $w$CDM model. Taking all quoted uncertainties at face value, combining either the SNLS SNe or distance ladder measurements with CMB data leads to a preference for $w<-1$ at $2\sigma$ \citep[95 per cent confidence limits of $w=-1.13^{+0.13}_{-0.14}$, and $w=-1.24^{+0.18}_{-0.19}$, respectively;][]{planckparams:prep}. If we consider two-dimensional contour plots from the triplet of parameters $\{w,\Omega_m,H_0\}$, a value of $w\simeq-1.2$, with $\Omega_m\simeq0.26$ and $H_0\simeq73~$km$~$s$^{-1}~$Mpc$^{-1}$, appears to effectively relieve the tension in the combined data set (Figure 3). For both the CMB and LSS, $w$ and $\Omega_m$ are positively correlated, and $w$ and $H_0$ are anti-correlated. Thus, allowing $w<-1$ both increases $H_0$, improving agreement with the distance ladder measurements, and decreases $\Omega_m$, improving agreement with the SNe.

\begin{figure}
	\centering
	\includegraphics[]{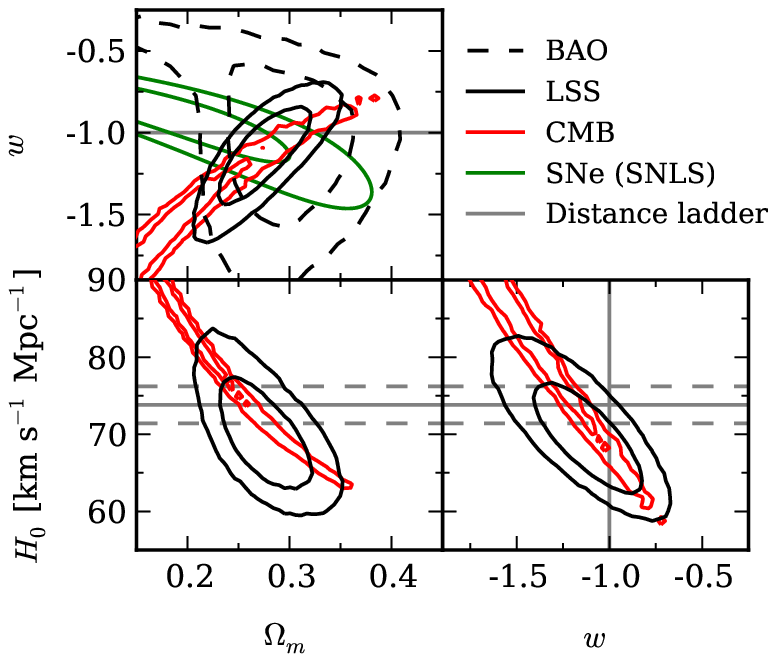}
	\caption{Marginalized two-dimensional constraints from the triplet of parameters $\{w,\Omega_m,H_0\}$ for the $w$CDM model. We compare results from the BAO and LSS clustering analyses in this work with the latest CMB constraints \protect\citep{planckparams:prep}, the SNLS SNe compilation \protect\citep{conley/etal:2011}, and local distance ladder $H_0$ measurements \protect\citep{riess/etal:2011}. Decreasing $w$ below $-1$ appears to relieve the $\Lambda$CDM tension between these data sets when these three parameters are considered -- but see Figure 4.}
\end{figure}

The situation changes somewhat when we consider LSS growth rate constraints on $\sigma_8$ with $w$ free. In Figure 4, we show constraints in the $w-\sigma_8$ plane. In order to include distance ladder and SNe constraints in this comparison, we used the publicly available \emph{Planck} CMB plus SNLS MCMC chain that was importance sampled with the \cite{riess/etal:2011} $H_0$ prior, having established that there is statistical agreement between the CMB, SNe and distance ladder data in the $w$CDM model (Figure 3).

Decreasing $w$ below $-1$ quickly leads to tension between the LSS and CMB determinations of $\sigma_8$, which are in good agreement for $\Lambda$CDM -- see Table 2. Consequently, when the constraints on $\sigma_8$ from the growth rate measurements are considered, allowing $w<-1$ appears less effective at relieving tension in the combined CMB, LSS, distance ladder and SNe data set.

It is worth examining why the CMB and LSS contours in Figure 4 are roughly orthogonal. For the flat $w$CDM model, the constraints on $w$, $\Omega_m$ and $H_0$ from the CMB are driven by the requirements that the angular diameter distance to last scattering, and the physical matter density, $\Omega_mh^2$, remain roughly fixed. This is achieved via the contours shown in Figure 3, and for $w<-1$ leads to a universe with a lower fractional matter density at the present time than inferred for $w=-1$. The dominant effect for $\sigma_8$ is an increase in $H_0$, which means $\sigma_8$ corresponds to rms mass fluctuations in smaller spheres (by definition, $\sigma_8$ is the rms fluctuation in spheres of radius $8~h^{-1}~$Mpc). The CMB prediction for $\sigma_8$ is therefore higher for $w<-1$ than for $\Lambda$CDM.

The LSS growth rate measurements probe $\sigma_8$ directly though $f\sigma_8$, which depends on the growing mode $\delta$ as 
\be
\begin{split}
f(z)\sigma_8(z)&=\frac{d\ln\delta}{d\ln a}\frac{\delta(z)}{\delta(0)}\sigma_8\\
&=-\frac{1+z}{\delta(0)}\frac{d\delta}{dz}\sigma_8.
\end{split}
\ee
If $w<-1$, the expansion history measured by current BAO and AP data can be recovered by decreasing $\Omega_m$, as for the CMB (Figure 3). At $z\sim0.6$, where the growth rate constraints are strongest, the dominant effect is an increase in $(-d\delta/dz)$, and, consequently, a decrease in $\sigma_8$ to balance the right-hand side of equation (8).

We can quantitatively assess the contribution of the growth rate constraints by importance sampling \citep[e.g.][]{lewis/bridle:2002} the CMB plus BAO chain supplied by the \emph{Planck} collaboration. We modify the chain sample weights using the ratio of the LSS likelihood adopted in this work to the BAO-only likelihood adopted by \cite{planckparams:prep}, which used the BAO constraints from 6dFGS, SDSS and BOSS (CMASS). We find a constraint of
\be
\begin{split}
w=-&1.03^{+0.06\,+0.12}_{-0.06\,-0.12}\\&\textrm{ (CMB+LSS; 68 and 95 per cent confidence)},
\end{split}
\ee
which is around 30 per cent tighter than in the original chain ($w=-1.08^{+0.09\,+0.18}_{-0.09\,-0.21}$). There are correspondingly tight constraints on other parameters, which are in good agreement with the $\Lambda$CDM values in Table 2: we find $H_0=(68.9\pm1.7)~$km$~$s$^{-1}~$Mpc$^{-1}$, $\Omega_m=0.297\pm0.015$ and $\sigma_8=0.829\pm0.018$. We expect importance sampling the CMB plus BAO chain to be a reasonable approximation to running a full chain with the LSS constraints included, since the $w$CDM degeneracies in the CMB data are already largely broken with the BAO data.

\begin{figure}
	\centering
	\includegraphics[]{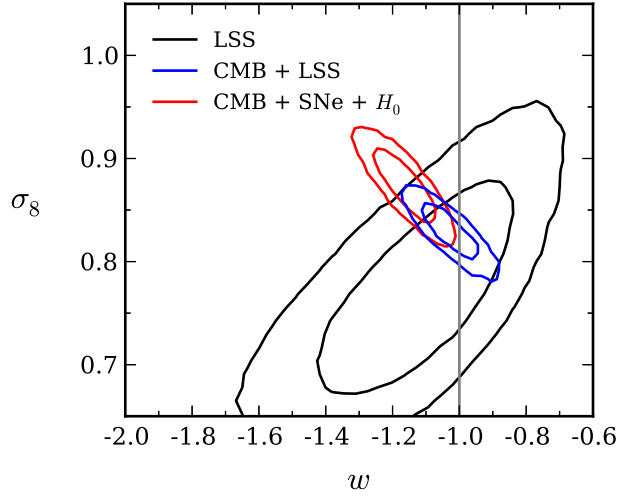}
	\caption{Marginalized $\sigma_8$ and $w$ constraints (68.3 and 95.5 per cent confidence) in the $w$CDM model. While tension in the combined CMB, SNe and distance ladder data set is effectively relieved by allowing $w<-1$ (Figure 3), the resulting $\sigma_8$ constraint is in moderate tension with the large-scale structure growth rate constraints from redshift-space distortion measurements. Note that the CMB and LSS constraints on $\sigma_8$ are in good agreement for the $\Lambda$CDM model (Table 2); the combined CMB and LSS constraint is $w=-1.03\pm0.06$.}
\end{figure}

\subsection{Comparison with cluster counts and weak lensing shear correlation measurements}

Incorporating growth rate constraints from redshift-space distortions also allows us to compare the LSS constraints in the $\sigma_8-\Omega_m$ plane to other low-redshift probes of the amplitude of matter fluctuations, including counts of galaxy clusters and galaxy weak gravitational lensing measurements. This comparison is made in Figure 5 assuming a $\Lambda$CDM model, using a pair of recent constraints. We have plotted contours corresponding to $\sigma_8(\Omega_m/0.27)^{0.6}=0.79\pm0.03$ from the weak lensing shear correlation function analysis of data from the Canada-France Hawaii Telescope Lensing Survey \citep[CFHTLens;][]{kilbinger/etal:2012}, and $\sigma_8(\Omega_m/0.29)^{0.322}=0.775\pm0.010$ from a cosmological analysis of clusters selected by the thermal Sunyaev Zel'dovich (SZ) effect using \emph{Planck} \citep{planckclusters:prep}. We do not see any degree of tension between the LSS clustering and other data sets here, although there is clearly some tension between the CMB power spectrum data and the weak lensing and cluster count constraints \citep[see also Sections 5.5.2 and 5.5.3 of][]{planckparams:prep}.

There is minimal correlation between $\sigma_8$ and $\Omega_m$ from current LSS data; making a more meaningful comparison using LSS clustering essentially requires tighter constraints on the growth rate to better measure $\sigma_8$.

\begin{figure}
	\centering
	\includegraphics[]{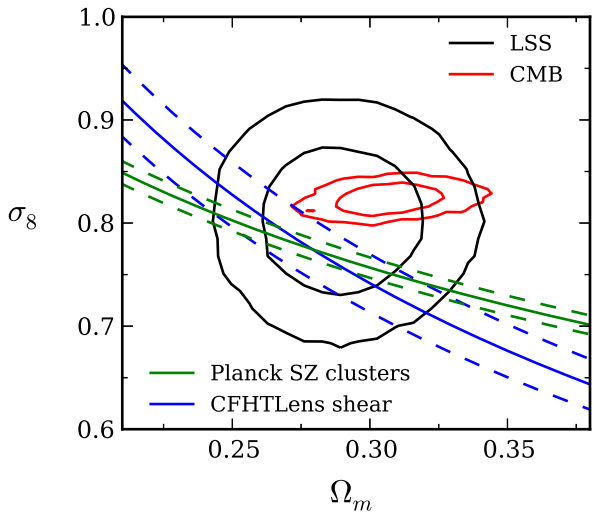}
	\caption{Comparison of $\Lambda$CDM constraints (68.3 and 95.5 per cent confidence) in the $\sigma_8-\Omega_m$ plane for CMB and several low-redshift data sets. While growth rate information from redshift-space distortions allows LSS clustering measurements to constrain $\sigma_8$, current precision is too low to meaningfully inform the comparison of CMB constraints with those from (for instance) cluster abundance \protect\citep{planckclusters:prep} or weak lensing shear correlations \protect\citep{kilbinger/etal:2012}. }
\end{figure}

\subsection{Future data}

The Ly-$\alpha$ data are a powerful complement to the galaxy clustering measurements at lower redshift. Without the BOSS Ly-$\alpha$ point, errors on $\Omega_m$ are more-than doubled for the BAO-only $\Lambda$CDM model, and tripled for $w$CDM. Constraints on $w$ itself are also greatly degraded, such that $w$ is almost unconstrained from below. When additional LSS constraints are included, the relative importance of the Ly-$\alpha$ constraint is diminished; removing the Ly-$\alpha$ constrain degrades the LSS $w$ constraint by around 15 per cent, from $-1.14\pm0.19$ to $-1.18\pm0.22$. Given that the Ly-$\alpha$ BAO constraints rely on analysis methodology less mature than used for the galaxy clustering measurements, it is encouraging that the shift in mean $w$ value from removing the Ly-$\alpha$ constraint is small compared to uncertainties.

As mentioned in Section 1, the quality and quantity of BAO position and LSS clustering data will improve considerably in coming months and years. The BOSS survey is expected to provide spectroscopic detections of roughly three times more galaxies than in the DR9 release, and around 50 per cent more quasars, leading to significant improvements over the constraints used in our analysis, particularly for the Ly-$\alpha$ BAO measurements \citep{boss:prep}. Various upcoming surveys are targeting BAO in the $z>1$ universe, including the Dark Energy Survey\footnote{http://www.darkenergysurvey.org/} \citep{des:prep}, MS-DESI, HETDEX\footnote{http://hetdex.org/} \citep{hill/etal:2008} and, looking further ahead, \emph{WFIRST}\footnote{http://wfirst.gsfc.nasa.gov/science/de/} and Euclid\footnote{http://sci.esa.int/science-e/www/area/index.cfm?fareaid=102} \citep{euclid:prep}. Many of these experiments will also attempt to constrain DE in other ways, using type IA SNe, galaxy weak lensing and cluster abundance, for example.

Telescopes including the Canadian HI Intensity Mapping Experiment\footnote{http://chime.phas.ubc.ca/} (CHIME), currently under construction in western Canada, plan to measure high-redshift BAO using HI intensity fluctuations. If the Galactic synchrotron foreground can be removed, CHIME has the potential to cheaply and quickly yield very competitive DE constraints from a diffuse tracer of LSS.

We also note that BAO constraints at low redshift ($z\lesssim0.2$), from surveys such as WALLABY\footnote{http://www.atnf.csiro.au/research/WALLABY/} and TAIPAN, would significantly improve on constraints from the 6dFGS \citep{beutler/etal:2011}. Amongst other things, these data would yield strong constraints on $H_0$ through an approach like that discussed in Section 5.1, when combined with high-redshift information.

We have highlighted the usefulness of growth rate constraints in both improving DE constraints and assessing consistency between data sets (Sections 5.2 and 5.3). There are also good prospects for future improvement in these measurements \citep[e.g.][]{weinberg/etal:prep}.

Our analysis of current LSS constraints demonstrates that we are already able to make meaningful comparisons between low-redshift data even without strong priors from the high-redshift universe probed by the CMB. It seems likely, if not inevitable, that future $\Lambda$CDM tensions between different cosmological probes will continue to arise as new data become available and constraints become more precise. Even if such tensions do not end up being ascribed to new physics, we must put ourselves in a position to make that assessment as robustly as possible. Examining different combinations of low-redshift data with and without CMB constraints should be a useful part of this process.

\section{Conclusions}

We have performed joint fits of cosmological parameters to current BAO position and LSS clustering measurements. We have shown that the BAO and LSS data are now of sufficiently high precision to make useful comparisons with other low-redshift cosmological probes in the virtual absence of CMB anisotropy constraints. We find that the BAO and LSS constraints are in good agreement with the latest CMB results from \emph{Planck} for the $\Lambda$CDM model, and mildly prefer a lower value of $H_0$, and higher value of $\Omega_m$, than some recent local distance ladder and type IA SNe measurements (Sections 5.1 and 5.2).

We note that the $\Lambda$CDM tension between \emph{Planck}, distance ladder and SNe data reported by \cite{planckparams:prep} appears to be effectively relieved by allowing $w<-1$, and that the CMB and LSS clustering data separately tolerate such behaviour. We show that the growth rate constraint on $\sigma_8$ from redshift-space distortions, is, however, in some tension with the combined CMB, distance ladder and SNe constraint in the $w$CDM model. Combining CMB, BAO and LSS data, including the growth rate information, we find $w=-1.03\pm0.06$; this constraint is around 30 per cent tighter than for CMB plus BAO position only, and completely consistent with $\Lambda$CDM.\\
\\
GA acknowledges support from a Canadian Institute for Theoretical Astrophysics (CITA) National Fellowship. This work was also supported by the Natural Sciences and Engineering Research Council of Canada (NSERC) and the Canadian Institute for Advanced Research (CIFAR). The authors would like to thank Chris Blake, Nicol\'as Busca, Chia-Hsun Chuang and Will Percival for clarification regarding the BAO and LSS measurements, Wendy Freedman for information relating to the distance ladder $H_0$ measurements, and Alex Conley for providing likelihood contours from the \cite{conley/etal:2011} SNLS analysis. We also thank the referee for useful comments and suggestions.

%\footnotesize{
% \bibliography{/Volumes/Data/Users/addisong/Documents/bibliography_files/act_trim,/Volumes/Data/Users/addisong/Documents/bibliography_files/wmap_jo,/Volumes/Data/Users/addisong/Documents/bibliography_files/wmap_supp,/Volumes/Data/Users/addisong/Documents/bibliography_files/graeme}
% }
 
 \footnotesize{
 
 }

\end{document}